 \documentclass[preprint2]{aastex}
\input{psfig}

\def\hi{\ifmmode {\rm H}\,{\sc i}~ \else H\,{\sc i}~\fi}

\def\chandra {{\it Chandra}~}

\def\gax    {${_>\atop^{\sim}}$}

\def\etal   {{\it et~al.}~}

\slugcomment{Submitted to ApJ Letters, \today}

\shorttitle{High redshift X-ray cluster}
\shortauthors{Mathur \& Williams}

\begin{document}


\title{\boldmath{\it Chandra} discovery of the intracluster medium around 
      UM425 at redshift 1.47}


\author{Smita Mathur and Rik J. Williams}
\affil{Astronomy Department, The Ohio State University, 140 West 18th Avenue, 
Columbus OH 43220}
\email{smita@astronomy.ohio-state.edu}



\begin{abstract}

We report on a discovery of a candidate cluster of galaxies at
redshift z=1.47 based on \chandra observations in the field of quasars
UM425 A \& B. We detect with high significance diffuse emission due
the intracluster hot gas around the quasar pair. This is the second
highest redshift cluster candidate after 3C294 at z=1.786. The diffuse
emission is elliptical in shape with about $17^{\prime\prime}$
extent. If indeed at z=1.47, this corresponds to a physical size of
140 h$_{70}^{-1}$ Kpc and 2--10 keV luminosity of $\sim 3\times
10^{43}$ erg s$^{-1}$.  The cluster is unlikely to be the long sought
gravitational lens invoked to explain unusual brightness of UM425 A
and the close quasar pair. Coexistence of the quasars with the cluster
suggests a link of activity to cluster environment. The unusual
brightness of UM425 A may then be due to a higher accretion rate. We
also comment briefly on the X-ray spectra of UM 425 A \& B which also
happen to be broad absorption line quasars. We argue that present
evidence suggests that the quasars are just a pair and not lensed
images of the same quasar.
\end{abstract}




\section{Introduction}

It is becoming increasingly clear the we live in a weird Universe,
filled with about 73\% dark energy, 23\% dark matter and only some 4\%
ordinary matter (Bennett \etal 2003, Spergel \etal 2003). One
expectation from this ``low density'' cosmological model is that large
scale structures formed early in time. A number of cluster surveys are
designed to find high redshift clusters of galaxies in optical and
radio wavelengths (e.g. Shectman \etal 1996, Kurk \etal 2001) to
provide independent cosmological constraints. In X-rays, clusters are
identified by the diffuse thermal emission from the intracluster gas,
which in fact accounts for most of their baryonic mass. A number of
X-ray surveys have been very successful in finding clusters,
determining their physical properties and using them as cosmological
tools (e.g. Vikhlinin \etal 1998, Vikhlinin \etal 2002a).

\chandra, because of its exquisite mirrors (Van Speybroeck \etal 1997) 
 and detectors (Garmire \etal, in preparation), started a new era of
 cluster research. The sub-arcsecond point spread function and low
 background of \chandra allowed detailed studies of low redshift X-ray
 bright clusters (e.g. McNamara \etal 2000, Fabian \etal 2000) and
 also led to the discovery of the highest redshift cluster candidate
 (at z =1.78, Fabian \etal 2001). While the clustering properties of
 low redshift clusters are used to determine cosmological parameters
 (e.g. Bahcall \etal 1999, Schuecher \etal 2001), just the number
 density of hot, high redshift clusters can also provide constraints
 to the matter density of the Universe (Donahue \etal 1998, Fabian
 \etal 2001). For example, the existence of z=1.78 cluster with
 temperature of $>3$ keV is inconsistent with a $\Omega_{m}=1.0$
 Universe (Fabian \etal 2001). Finding hot, X-ray clusters at high
 redshifts, therefore, becomes important.

Here we report a serendipitous discovery of a cluster candidate
 traced by hot intracluster medium around quasar pair UM425 A
\& B at z=1.47. This quasar pair, separated by $6.5^{\prime\prime}$, was
discovered by Meylan \& Djorgovski (1989) as a gravitational lens
candidate because of the unusually large luminosity of the brighter of
the pair and its relatively large redshift. Ever since, whether the
pair represents lensed or binary quasars has been a matter of
debate. To cause such a wide angle separation a massive cluster of
galaxies along the line sight would be required, but none was found
down to a limiting magnitude of $m_{\rm R}\sim 24$ (Courbin \etal
1995). UM425 A \& B also happen to be broad absorption line quasars
(BALQSOs). BALQSOs are extremely faint X-ray sources, and in
pre-\chandra era, X-ray spectroscopy of BALQSOs was practically
impossible (Mathur \etal 2000 and references there in). With \chandra,
many BALQSOs were detected in X-rays (Green \etal 2001, Gallagher
\etal 2002), but spectroscopy was feasible for only a handful of
them. The original objective behind our \chandra proposal was to study
the X-ray spectrum of UM425 A, taking advantage its unusual
brightness. While the main focus of this paper is the
unexpected discovery of diffuse intracluster medium, we also briefly
discuss the BALQSOs. A cosmological model with $H_0= 70$ km s$^{-1}$
Mpc$^{-1}$, $\Omega_m=0.3$ and $\Omega_{\lambda}=0.7$ is used throughout
the paper.

\section{Observation and Analysis}
\subsection{\chandra observation}
We observed UM425 with \chandra on 13 December 2001 for a exposure
time of 110 ksec. The advanced CCD imaging camera for spectroscopy
(ACIS-S, C. R. Canizares \etal in preparation) was used at the nominal
aimpoint. The data were reduced in a standard manner using \chandra
interactive analysis software (CIAO version 2.3, Elvis, M. \etal in
preparation) and following the thread for imaging analysis
\footnote{see: http://asc.harvard.edu/ciao/threads/index.html}. Figure
2 shows the soft band (0.3--3 keV) image of the field. The quasars
UM425 A \& B are clearly detected with $4675\pm69$ and $20\pm5$ net,
background subtracted counts each in the broad band (0.3--8 keV). In
addition, faint extended emission is clearly seen extending to
west-southwest of UM425 A. Contours of 2,3,4,5 and 6 times the
background are overlayed on the image showing the extent and
significance of the diffuse emission. Figure 2 shows the adaptively
smoothed image of the field in the soft band created using CIAO tool
{\bf csmooth} and used to generate the contours mentioned above. We
used a Gaussian smoothing function with a threshold of 2$\sigma$. The
smooth image was then divided by the smoothed exposure map. The
resulting image is displayed in figure 2 and clearly shows the diffuse
emission.

\begin{figure}[ht]
\psfig{figure=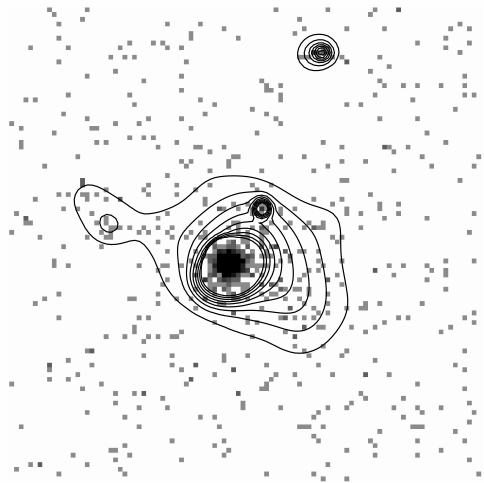,height=3in}
\caption{The soft band (0.3--3 keV) image of the UM425 field
 overlayed with contours of 2 to 6 times background level. The
 contours were generated from the smooth image shown in figure 2. In
 addition to the two point sources UM425 A \& B, the diffuse emission
 is also evident. The bright source about $25^{\prime\prime}$
 north-north-west of UM425 is a foreground galaxy at z=0.1265 (Meylan
 \& Djorgovski 1989)}
\end{figure}

\subsection{The extended diffuse emission}

The diffuse emission is somewhat elliptical in shape and extends to
about $10^{\prime\prime}$ west-southwest of UM425 A with total extent
of about $17^{\prime\prime}$. This corresponds to about 140 kpc at
z=1.47. Because of the presence of the strong point source (UM425 A),
it is difficult to determine the exact number of counts from the
diffuse emission. First, we extracted counts from a smaller region
well outside the area occupied by the point sources. Then we scaled
these counts by the area of the entire diffuse emission as shown in
figures 1 and 2. In this way we estimate 130 counts from the diffuse
emission. Background counts from a region of same size are estimated
to be about 38. With $98\pm13$ net counts, the diffuse emission is
thus highly significant at $7.6\sigma$ level. 

As another method of estimating the total counts from the diffuse
emission, we extracted all counts from an annular region surrounding
UM425 A.  The inner $2^{\prime\prime}$ core of the bright point source
was excluded, and the extraction region extended out to a radius of 12
arcsec, completely covering the apparent extent of the diffuse
emission.  We then used the
ChaRT\footnote{http://asc.harvard.edu/chart/} and
MARX\footnote{http://asc.harvard.edu/chart/threads/marx/} tools to
simulate the point spread function (PSF) of UM425 A, and used the same
region described above to extract the counts from the outer wings of
the simulated PSF.  There were $281\pm22$ counts in the image
extraction region, $157\pm13$ in the simulated PSF region, and
$20\pm5$ in UM425 B, resulting in $104\pm26$ net counts from the
diffuse emission, consistent with the estimate above.

For spectral analysis, we used only the counts from the small 
extraction region defined above, which is about half the size of the
total area of the diffuse emission.  With just 46 net counts, spectral
shape cannot be determined accurately. Nevertheless, we performed
spectral analysis to obtain rough estimate of the temperature of the
diffuse plasma using {\it Sherpa} spectral fitting package within
CIAO. Using Raymond-Smith thermal plasma model, and fixing the
abundance to 0.1 solar and foreground absorbing column to Galactic
($4.1\times 10^{20}$ atoms cm$^{-2}$), we find the best fit
temperature to be 2.1 keV at the redshift of the cluster, assumed to
be z=1.47. This estimate is highly uncertain, with lower limit on the
rest frame temperature being $kT$ \gax $0.8$ keV with 90\% confidence;
upper limit is unconstrained. Assuming the rest frame temperature of
$kT= 2.1$ keV, the observed unabsorbed flux is $f_X(0.1-2.5 keV)=
4.6\times 10^{-15}$ and $f_X(2-10 keV)= 4.6\times 10^{-16}$ erg
s$^{-1}$ cm$^{-2}$.

\begin{figure}[ht]
\psfig{figure=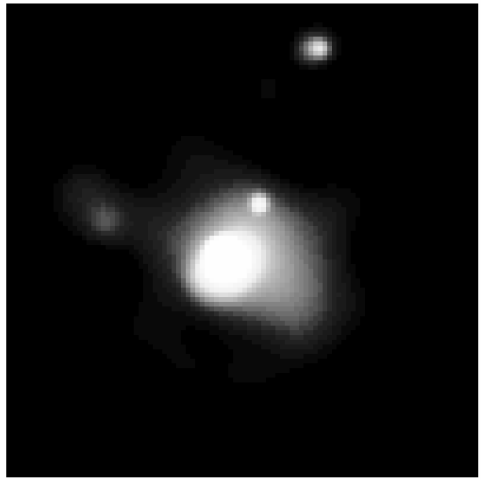,height=3in}
\caption{The adaptively smoothed, exposure corrected, soft band image
 of the UM425 field. In addition to the quasars UM425 A \& B, the
 diffuse cluster emission is clearly seen in this display.}
\end{figure}

\subsection{The BALQSOs}

Because BALQSOs are believed to be highly absorbed sources, we used
absorbed power-law models to fit their X-ray spectra. The presence of
excess soft photons, however, often points to partial covering of the
source by absorbing material (see Grupe, Mathur \& Elvis 2003 for detailed
procedure of spectral analysis that we follow). For UM425 A, the best
fit spectrum ($\chi^2=82.1$ for 112 degrees of freedom) yields
following parameters: power-law slope $\alpha=0.93\pm0.08$
($f_{\nu}\propto \nu^{-\alpha}$), column density at the source
N$_H=3.1\pm0.7\times 10^{22}$ cm$^{-2}$ and the covering fraction
$f=0.74\pm0.04$. This spectrum, and those of other BALQSOs, show that
strong absorption is the main cause of their X-ray faintness (Grupe 
\etal 2003, Green \etal 2001, Gallagher \etal 2001).

UM425 B has too few counts for accurate spectral analysis, but we can
test whether the UM425 A spectrum provides adequate description of
UM425 B data. Keeping all other parameters constant, and allowing only
normalization to be the free parameter, the fit is not good with too
few counts around a keV compared to the model. If we allow both
normalization and absorbing column density to vary, then the resulting
N$_H=5\times 10^{23}$ cm$^{-2}$, over an order of magnitude higher
than that in the A component. Alternatively, keeping the N$_H$
constant as in component A, but allowing the covering fraction to
vary, leads to $f$ reaching the hard limit of $f=1$. Thus the spectra
of UM425 A \& B are found to be inconsistent.

\section{Discussion}

 \chandra observations of UM425 field led to the discovery of hot
 intracluster medium around the quasar pair. High-z clusters are rare,
 and so are the large separation quasar pairs; chance association of
 these two uncommon phenomena is unlikely. Is the X-ray cluster at the
 redshift of the quasars, hosting the pair, or is it the foreground
 lens? Aldcroft \etal (2003) have shown that to produce the large
 observed separation between the quasar pair, the minimum cluster
 temperature will have to be $kT=1.5 keV$, with luminosity
 $L_X(0.1-2.4 keV) \approx 1.5 \times 10^{43}$ erg s$^{-1}$ at z=0.5
 resulting in observed flux of $f_X(0.1-2.5 keV) \approx 2.3 \times
 10^{-14}$ erg s$^{-1}$ cm$^{-2}$. Given the uncertainty, we cannot
 rule out the temperature of 1.5 keV at z=0.5 ($\S 2.2$). However, if
 the observed cluster has this temperature at z=0.5, the resulting
 flux is $f_X(0.1-2.4 keV)= 4.2\times 10^{-15}$ and $f_X(2-10 keV)=
 5.6\times 10^{-16}$ erg s$^{-1}$ cm$^{-2}$. Thus the observed X-ray
 flux is over five times fainter than the minimum flux estimated for
 the foreground lens, so the X-ray cluster is unlikely to be the lower
 redshift lensing cluster. UM425 field has been a target of cluster
 searches because the quasar pair was thought to be lensed and because
 there is a small overdensity of galaxies in the field, suggestive of
 a foreground cluster (Meylan \& Djorgovski 1989, Corbin \etal
 1995). However, no such cluster is found. This also suggests that the
 X-ray cluster reported here is at a much higher redshift. Deep
 imaging and spectroscopic studies will be required to confirm the
 redshift of the X-ray cluster, but it is highly likely
 to be at the redshift of the quasars, hosting the pair. This is
 exciting, because only one other cluster at a higher redshift is
 known (Fabian \etal 2001)

Assuming the cluster to be at z=1.47, we find its size to be 140 kpc
and luminosity $L_X(0.3-10 keV)=6.2\times 10^{43}$ erg s$^{-1}$ and
$L_X(2-10 keV)=3\times 10^{43}$ erg s$^{-1}$ for temperature kT=2.1
keV. These values of temperature and luminosity lie right on the
L$_X$--T relation for low redshift groups and cluster (Mulchaey \&
Zabludoff 1998), which is remarkable given the large uncertainty
determination of both the quantities and possible evolution of the
L$_X$--T relation with redshift (Vikhlinin \etal 2002b). The luminosity
is at the lower end of that found for clusters. The observed size of
the diffuse emission, however, is small, more like that of a
group. The faint diffuse emission that we detect must trace the more
luminous intracluster gas from the cluster core. Our estimates of
temperature and luminosity may then be considered as lower
limits. Given the serendipitous nature of this discovery, we cannot
determine the space density of such hot high redshift clusters. The
presence of a few such clusters on the sky at redshifts of about 1.5
and higher is consistent with the currently popular cosmological model
with $\Omega_m=0.3$ and $\Omega_{\lambda}=0.7$, but we cannot rule out
the $\Omega_m=1.0$ model based on just this observation with poor
temperature constraint (c.f. Fabian \etal 2001, who find that the
maximum number of collapsed objects with temperature larger than 10
keV to be about 100 for redshifts higher than z=1.8).

The association of the X-ray cluster with the quasars also implies
that they are just a pair of quasars, not lensed images of the same
quasar.  This interpretation is also supported by the different X-ray
spectra of the two objects (though, in principle, slightly different
sightlines to the nuclear X-ray source can explain the observed
difference). We can also explain the unusual brightness of UM425 A
without invoking lens magnification.  Association of UM425 A \& B with
the X-ray cluster, and observations by Martini \etal (2002) who found
high fraction of AGNs in cluster Abell 2104, suggests that a cluster
environment is conducive for triggering of AGN activity.  UM425 A is
about two magnitudes brighter than other quasars at similar redshifts
(Aldcroft \etal 2003). Czerny \etal (1997) have shown that most
quasars usually radiate at $\sim 0.01-0.2$ of their Eddington
luminosity. If UM425 A is radiating at close to Eddington limit, it
can easily have an order of magnitude larger luminosity for similar
black hole mass and radiative efficeincy. Such a high accretion rate
relative to Eddington may be related to the recent triggering of the
quasar (Mathur 2000) and supports the hypothesis that BALQSOs
represent an early phase in quasar lifetimes (Canalizo \& Stockton
2001, Fabian 1999, Becker \etal 2000). If the fraction of quasars with
unusually high luminosity is small, it may imply that the high
accretion phase does not last long. The faint B component was only
discovered because of its close proximity to UM425 A.

To summarize, our \chandra observations led to two interesting
results. First, the discovery of a cluster candidate with second
highest redshift and second, evidence suggesting a connection between
cluster environment and quasar activity and supporting evolutionary
hypothesis to understand the BALQSO phenomenon.

\acknowledgments

We are grateful to the entire \chandra and NASA community for a superb
mission. We thank David Weinberg, Brian McNamara \& Brad Peterson for useful
discussions and Tom Aldcroft \& Paul Green for their contribution to
our \chandra proposal.


\clearpage




\end{document}